\documentclass[aps,twocolumn,groupedaddress,nofootinbib,pre]{revtex4-1}
\usepackage[pdftex]{graphicx}
\usepackage{latexsym,amsmath,verbatim}
\usepackage{xcolor}
\usepackage{lipsum}
\usepackage[english]{babel}
\newcommand{\be}{\begin{equation}}
\newcommand{\ee}{\end{equation}}

\begin{document}
\title{Analytical study of quality-biased competition dynamics for memes in social media}

\author{Daniele Notarmuzi}
\affiliation{Dipartimento di Fisica, 
Sapienza Universit\`a di Roma, P. le A. Moro 2, I-00185 Roma, Italy}

\author{Claudio Castellano}
\affiliation{Istituto dei Sistemi Complessi (ISC-CNR), Via dei Taurini 19, 
I-00185 Roma, Italy}
\affiliation{Dipartimento di Fisica, 
Sapienza Universit\`a di Roma, P. le A. Moro 2, I-00185 Roma, Italy}


\begin{abstract}
The spreading of news, memes and other pieces of information 
occurring via online social platforms has a strong and growing impact 
on our modern societies, with enormous consequences, that may be 
beneficial but also catastrophic.
In this work we consider a recently introduced
model for information diffusion in social media taking explicitly 
into account the competition of a large number of items of diverse
quality. We map the meme dynamics onto a one-dimensional 
diffusion process that we solve analytically, deriving the lifetime
and popularity distributions of individual memes. We also present a
mean-field type of approach that reproduces the average stationary
properties of the dynamics. In this way we understand and control
the role of the different ingredients of the model, opening the
path for the inclusion of additional, more realistic, features.
\end{abstract}

\maketitle

\section{Introduction}
Understanding how information spreads in social media is a topic of 
uttermost interest, as it is fundamental for devising strategies 
aimed at fostering the diffusion of beneficial information or contrasting the
dangerous spread of fake news~\cite{Howell2013,DelVicario554,Vosoughi1146}.
Activity in this area has boomed in recent 
years~\cite{Kwak2010,Lerman2010,GonzalezBailon2011,Bakshy2011,Banos2013,Cheng2014,Nishi2016,Pramanik2017,Wegrzycki2017}.
From the point of view of statistical physics, information spreading is a
prominent example of a collective macroscopic phenomenon emerging
in a self-organized manner from the spontaneous activity of a large number
of individual elements~\cite{Castellano2009,Buchanan2007}.
The investigation of information spreading is particularly challenging 
both from an empirical point of view and from a theoretical one.
The existence of many different social media platforms, each characterized
by different features often changing over time, provides a wealth of data
but leaves the issues of universality and reproducibility wide open.
From the modeling point of view, the identification of a limited number 
of relevant mechanisms and crucial observable quantities is highly nontrivial.

The topology of the interaction pattern among users in online social media,
which is usually very heterogeneous, is one of the ingredients usually taken
into account.
Another fundamental factor affecting the way news, memes or rumors are
diffused is information overload. When online, individuals are hit 
by a steady and overwhelming
flow of messages; the finite attention and limited memory strongly 
influence what information is propagated further and how. This results
in a competition among a large number of items diffusing 
simultaneously, which is a key ingredient of many models for information 
spreading~\cite{Wu2007,Weng2012,Gleeson2014,Gleeson2016}.
A third ingredient that plays a role in determining the fate of messages
in online media is the variability of the ``quality'' of the item: 
some pieces of information may be intrinsically more appealing and thus more 
likely to be shared by online users.
A very recent work by Qiu et al.~\cite{Qiu2017} considered together these 
three elements to study the interplay of an heterogeneous quality 
distribution and  information overload in online social media (with
particular reference to Twitter), with the goal of investigating 
whether a good tradeoff between discriminative power and quality diversity 
is possible.

Although highly stilized, the model for meme dynamics introduced 
in Ref.~\cite{Qiu2017}
contains several relevant ingredients of the real phenomenon and in particular
the original element that the competition among different memes favors those
having a higher intrinsic quality. For this reason we call it the quality-biased
competition (QBC) model.
In this paper we study the QBC dynamics in detail, by considering some 
carefully devised simplifications, which make possible an analytical 
treatment providing 
explicit formulas for the behavior of the main observables. 
In this way we achieve a full understanding of the model phenomenology 
and of its dependence on the value of the different parameters.

\section{The QBC model}
We consider the model for meme spreading introduced in Ref.~\cite{Qiu2017}.
$N_u$ agents (or users), each of them equipped with a memory 
containing at most $\alpha$ memes, are the nodes of a static network.
Memories are ordered lists from $\alpha$ to 1.
At each time step an individual is selected uniformly at random and 
transmits a meme to all her neighbors. 
With probability $1-\mu$, the transmitted meme is an existing one,
taken from the agent memory; otherwise, with probability $\mu$,
a new meme is created. 
In both cases, the transmitted meme is put at the top (position $\alpha$) 
of the memory 
of the agents involved (both the transmitter and the receivers)
shifting all other memes downward.
Each meme is attributed randomly, upon its creation, a fitness
value $f_i$ between 0 and 1, a proxy of its quality.
When a user selects an old meme for transmission, the probability to 
select meme $i$ is proportional to $f_i$.
In this way high fitness increases the chance of the meme to be spread.
Apart from this bias, the dynamics can be
seen as the competition among many susceptible-infected spreading
processes in a metapopulation framework~\cite{PastorSatorras2015}.

From the initial configuration with all empty memories, memes are
introduced and copied until some of the memories fill up.  When all
slots in a memory are occupied and a new meme must enter, the item in
the last position is removed and forgotten by the agent.
Memories thus work according to a
"first-in first-out" rule, mimicking what happens on users feeds 
of some social networks, such as Twitter.
After an initial transient, a steady state is reached where all 
$N_u \alpha$ memory slots in the system are occupied. 
Memes are continuously created, diffuse over the network 
and get eventually extinct.
Quantities characterizing the dynamics of a meme are 
its lifetime, i.e., the time passed between the creation of a meme
and its extinction, and its popularity, defined as the total number
of times the meme is transmitted, throughout its lifetime, 
from an agent to one of her neighbors.

\section{Robustness with respect to the topology}
\begin{figure*}
\includegraphics[width=0.49\textwidth]{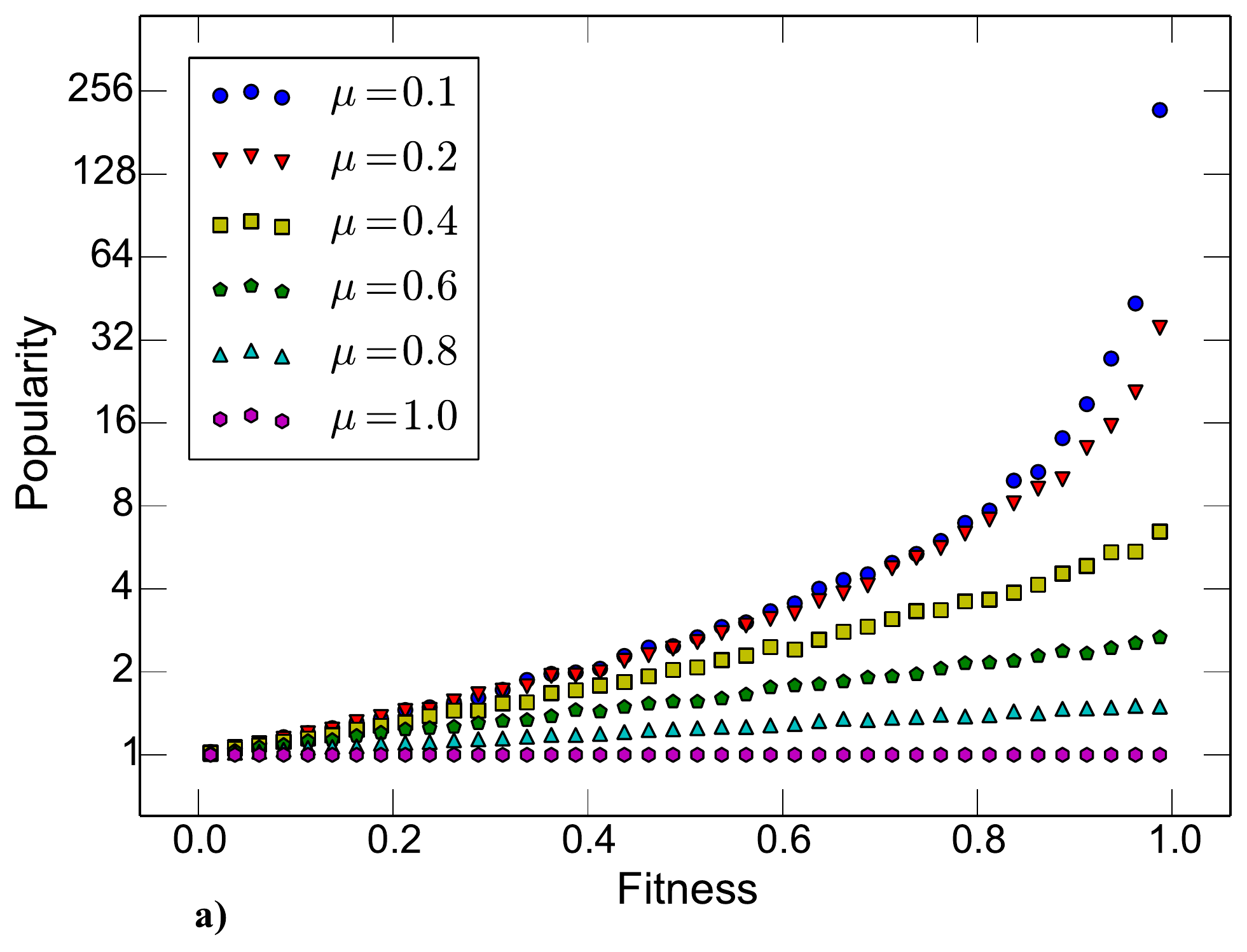}
\includegraphics[width=0.49\textwidth]{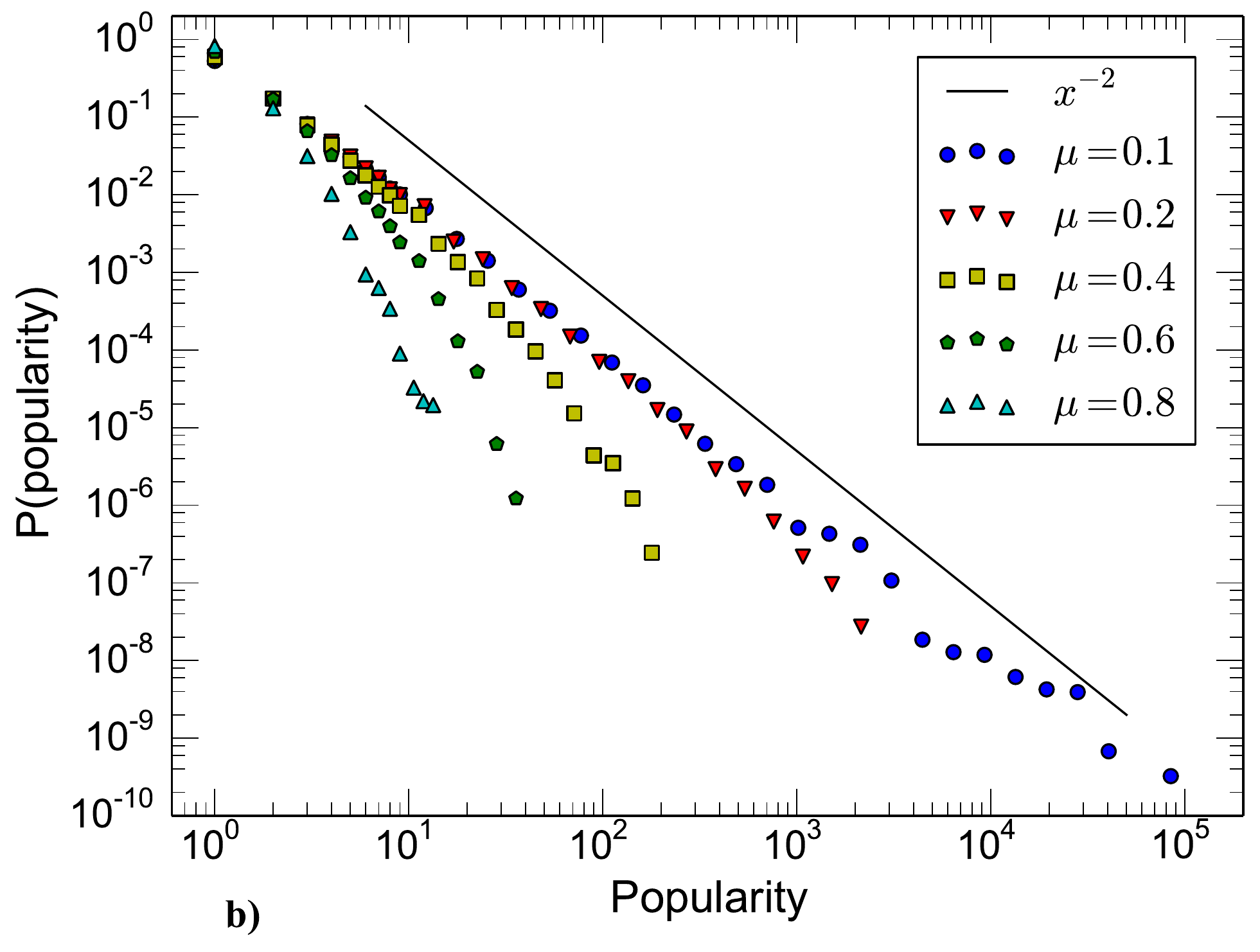}
\includegraphics[width=0.49\textwidth]{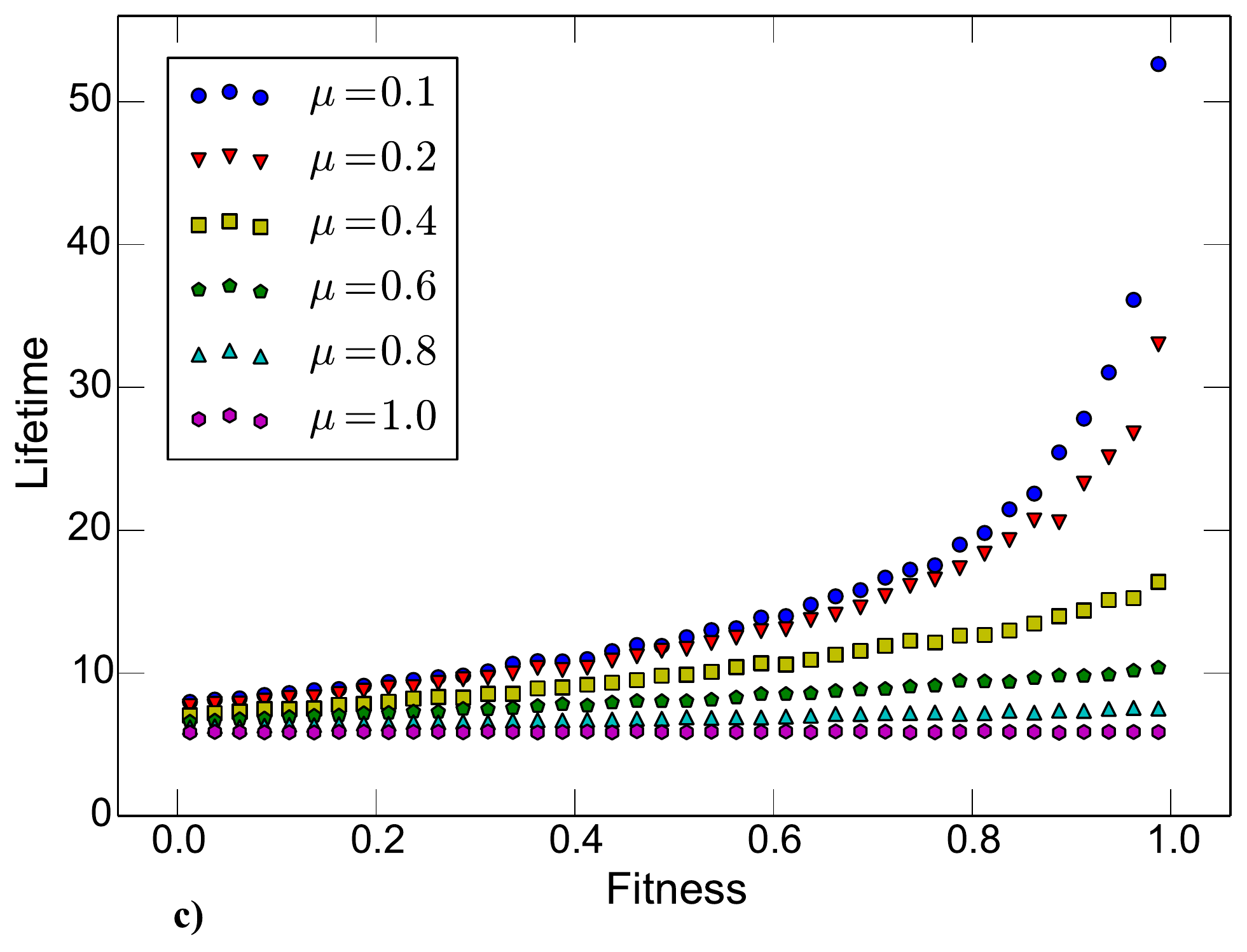}
\includegraphics[width=0.49\textwidth]{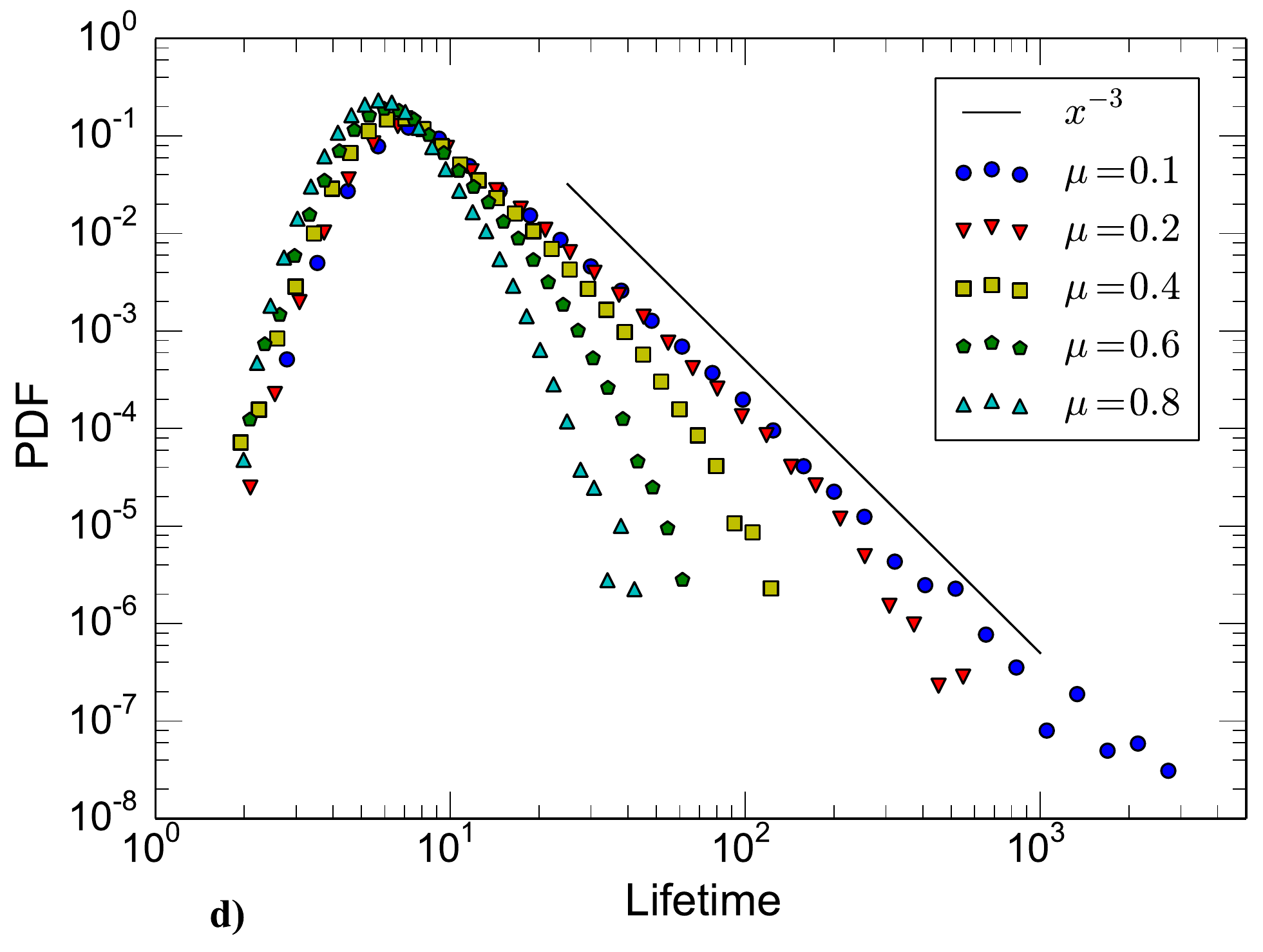}
\caption{a) Average popularity as a function of fitness in the QBC model
  on an annealed random regular graph with degree distribution
  $P(k)=\delta_{k,1}$ for different values of $\mu$ with fixed
  $\alpha=10$ and $N_u=10^3$. Averages are performed over $10^5$ memes. 
  b) Popularity probabilities for the same system.
  c) Average lifetime for the same system.
  d) Lifetime PDF for the same system.
}
\label{fig_annealed}
\end{figure*}

We first check how much the model phenomenology depends 
on details of the interaction pattern, by performing 
numerical simulations on several types of network 
(see Supplementary Material, SM).
It turns out that the distributions of the main observables
are qualitatively robust with respect to changes
of the underlying network (see Fig.~\ref{fig_annealed}). 
Both distributions have broad power-law tails, cutoff 
exponentially over a scale growing when $\mu$, the rate 
of creation of new items, goes to zero.
The lifetime distribution also exhibits a peak for $l$ of the
order of $\alpha$, corresponding to the average time needed 
for a meme that is not shared to disappear from the memory of
the agent that created it. The average values of the popularity
and of the lifetime strongly grow with the fitness when
$\mu$ is small.
The effect of the parameter $\alpha$ is very weak (see SM).
The overall picture remains the same even if the contact pattern is
an annealed random regular graph where each node has a single connection.
This suggests that a mean-field approach, which effectively 
considers a regular annealed network as contact pattern, may provide 
an accurate description of the model dynamics.

\section{A microscopic approach}
We focus now on the behavior of an individual meme of fitness $f$.
We define as $0 \le N_{ij}(t) \le \alpha$ the position of meme $i$ in 
the memory of agent $j$ at time $t$: $N_{ij}=\alpha$ corresponds to the 
top position (a newly created or transmitted meme), while $N_{ij}=1$ 
means that the meme is about to be forgotten. If meme $i$ does not appear 
in the memory of agent $j$, then $N_{ij}=0$. 
We neglect the case in which an agent has more copies of the same meme in 
his feed.
The quantity $N_i(t)=\sum_{j}N_{ij}(t)$ cumulates
the positions of the meme in all users' feeds, thus providing information
about its overall diffusion.
For simplicity we assume that each user is in contact with a single 
randomly chosen other user and that, when with probability $\mu$ a user 
produces a new meme, she simply puts it on top of her memory, without 
immediately sharing it. For the same reason we assume that, when an existing
meme is selected for transmission, it is left in the original position
in the transmitter feed, without putting it at the top ot the memory.
We checked that both these assumption have negligible effects.
The quantity $N_i(t)$ performs over time a one dimensional random-walk
in the interval $[0,\alpha N_u]$. 
$N_i=0$ is an absorbing boundary condition (after
extinction a meme will never reappear) and $N_i=\alpha N_u$ is a 
semireflecting boundary (because of our approximation, if the meme 
is in the first position of all feeds, $N_i$ cannot grow further).
The initial condition is $N_i=\alpha$.
At each time step the elementary events are:
\be
N_i(t) = \left\{
\begin{array}{lcl}
n \rightarrow n + \alpha & & \textrm{with prob. $R_n$} \\
n \rightarrow n - 1      & & \textrm{with prob. $L_n$} \\
n \rightarrow n          & & \textrm{with prob. $S_n=1-R_n-S_n$}.
\end{array}
\right.
\ee

Apart from different expressions close to the boundaries (see SM for details),
the probabilities are: 
\be
R_n = (1-\mu)\frac{C_n}{N_u}\left(1-\frac{C_n}{N_u}\right)\frac{f}{\alpha},
\label{R_n}
\ee
and
\be
L_n = \frac{C_n}{N_u}.
\ee
where $f$ is the fitness of the considered meme and $C_n$ is 
the number of individuals possessing $i$ in their memory.

Eq.~(\ref{R_n}) is derived based on the consideration that
$N_i$ is increased by $\alpha$ if a transmission event takes place
(it happens with probability $1-\mu$), if meme $i$ is present
in the feed of the transmitting user (probability $C_n/N_u$) and 
not present in the feed of the receiver $\left(1-C_n/N_u \right)$ 
and if meme $i$ is selected for transmission among all memes
in the feed. This last event occurs with probability 
$f_i / \sum\limits_{j\in M_u}f_j$, which we approximate with $f/\alpha$.
$C_n$ is the number of individuals possessing $i$ in their memory, that
we approximate as 
\begin{equation}
  \label{eq:copies_Ci}
  C_{n} = \left \lfloor\frac{n + \alpha -1}{\alpha} \right\rfloor \; ,
\end{equation}
where $\lfloor x \rfloor$ represents the integer part (floor) of $x$.

With regard to $L_n$, the value of $N_i(t)$ decreases because the
insertion of a new meme in a user feed causes the downward shift of
all other memes.  The insertion occurs at each time step, irrespective
of whether the inserted meme is new or transmitted.  Hence
$L_n=\frac{C_n}{N_u}$, the likelihood that meme $i$ is present in the
involved memory.  From the expressions of the probabilities it is
immediately clear that nothing depends on $f$ and $\mu$ separately,
but only through the combination $\beta=(1-\mu)f$.  

We simulate
numerically this random walk description of meme dynamics.  In the SM
we show that the popularity and lifetime distributions obtained match
very closely those found for the original QBC model.

In order to make the analytical treatment easier,
we further simplify the random-walk description. 
In particular, we remove the 
floor function from Eq.~(\ref{eq:copies_Ci}), we set equal to $1$ the 
term $\left(1-C_n/N_u \right)$ in Eq.~(\ref{R_n}) and 
we introduce a numerical constant 
$\gamma=(\alpha + 1)\frac{\alpha N_u + \alpha - 1}{\alpha^2 N_u}$ 
in the denominator of Eq.~(\ref{eq:copies_Ci}). See the SM for the justification
of these modifications.
Again we numerically check the distributions generated by this simplified
random-walk description and find (see SM) that they are essentially equal
to those of the original QBC dynamics.

At this point we can write down the master equation
for the modified random walk, which reads
\begin{subequations}
  \begin{equation}
    P_n(t+\Delta t)=S_nP_n(t)+L_{n+1}P_{n+1}(t)+R_{n-\alpha}P_{n-\alpha}(t)
    \label{eq:me-original_n}
  \end{equation}
  \begin{equation}
    P_{\alpha N_u}(t+\Delta t)=(1-\mu)P_{\alpha N_u}(t)+
    \sum\limits_{j=0}^{j=\alpha}R_{\alpha N_u-j}P_{\alpha N_u-j}(t)
    \label{eq:me-orginal_N}
  \end{equation} 
  \label{eq:me-original}
\end{subequations}
where Equation~(\ref{eq:me-original_n}) holds for $n=0,1,...\ \alpha N_u-1$ 
provided one considers $R_{n-\alpha}=0$ for $n=0,1,...\ \alpha$ and 
$\Delta t = N_u^{-1}$.

By setting $x_n\equiv n/ (\gamma \alpha N_u)$ with $x_n$ ranging between $0$ 
and $1/\gamma$ and taking the thermodynamic limit $N_u \rightarrow \infty$,
from the master equation we obtain (see SM) the Fokker-Planck (FP) equation
for the probability $\rho(x,t)$ that the walker is in position $x$
at time $t$:
\begin{equation}
  \frac{\partial}{\partial t}\rho(x,t)=\frac{1-\beta}{\gamma \alpha}
  \frac{\partial}{\partial x}x\rho(x,t) +						
  \frac{1+\beta\alpha}{2 \gamma^2 \alpha^2 N_u}\frac{\partial^2}
       {\partial x^2}x\rho(x,t) \; .
       \label{eq:FP_complete}
\end{equation}
For large $N_u$ we have $\gamma=(\alpha+1)/\alpha$.

\subsection{Purely diffusive dynamics}
In the limit $\beta \rightarrow 1$ the drift term in Eq.~(\ref{eq:FP_complete}) vanishes. 
We are left with the FP equation of a purely diffusive stochastic process:
\begin{equation}
	\frac{\partial}{\partial t}\rho(x,t)=D_0\frac{\partial^2}{\partial x^2}x
\rho(x,t) \; ,
	\label{eq:FP_pure-diff}
\end{equation}
where
\begin{equation}
  D_0 = \frac{1+\alpha}{2 \gamma^2 \alpha^2 N_u} = \frac{1}{2 (1+\alpha) N_u},
\end{equation}
which differs from standard diffusion because of the space-dependent
diffusion coefficient.
The limit $\mu \rightarrow 0$ changes also the boundary
conditions: the boundary in $x=1/\gamma$ is \textit{semireflecting}
because $N_i(t)=N_u \alpha$ can decrease with probability $\mu$ or remain
unchanged, with probability $1-\mu$.
Thus in the case $\mu =0$ both the boundary condition in $x=0$
and in $x=\gamma^{-1}$ are absorbing: $\rho(0,t)=\rho(\gamma^{-1},t)=0$.
The initial condition is $\rho(x,t=0)=\delta(x-x_{\alpha})$
with $x_{\alpha} =1/ ( \gamma N_u )$. 

\medskip

It is possible to find the solution of this equation as an
eigenfunction expansion of the operator 
$\mathcal{L}_{FP} = D_0 \frac{\partial^2}{\partial x^2} x$ 
(see SM for details), obtaining:
\begin{equation}
	\rho(x,t) = \frac{\pi^2 \gamma }{2}  \sqrt{\frac{x_{\alpha}}{x}} 
\sum_{n=1}^{\infty} n J_1(\pi n\sqrt{\gamma x_{\alpha}}) J_1(\pi n \sqrt{\gamma x})
 e^{-t/\tau_n}  \; ,
	\label{eq:solution_complete_pure-diff}
\end{equation}
where $J_1(z)$ is a Bessel function of the first kind.
The characteristic time scale of each eigenfunction is
\begin{equation}
  \tau_n = \frac{8\gamma\alpha^2 N_u}{(1+\alpha)j_{1,n}^2} \simeq
  \frac{8\alpha N_u}{\pi^2 n^2} .
  \label{eq:tau_n}
\end{equation}
where the $j_{1,n}$, the zeros of $J_1(z)$, are approximated as 
$j_{1,n}=\pi n$.
Using this expression, it is possible to compute (see SM for details)
the survival probability in the limit $\mu \rightarrow 0$, 
which turns out to be
\begin{equation}
	S(t) \simeq \left\{
\begin{array}{lcl}
1 & &  t \ll \alpha \\
\alpha t^{-1} e^{-t/\tau} & & t \gg \alpha
\end{array}
\right.
\end{equation}
where $\tau=8 \alpha N_u/\pi^2$ is $\tau_1$ 
after the approximation $j_{1,n} \rightarrow \pi n$ is made.
Based on this result the lifetime distribution can be
computerd (see SM).
In the limit of large $N_u$, i.e., diverging $\tau$, it reads
\begin{equation}
  F(l) \!= \!-\left. \frac{d S}{dt} \right|_{t=l} \!\simeq \!
\left\{ \!
\begin{array}{lr}
\! 0  & l \ll \alpha \\
\alpha l^{-2} & l \gg \alpha
\end{array}
\right.
\label{eq:F(l)_cut}
\end{equation} 

\begin{figure}
\includegraphics[width=\columnwidth]{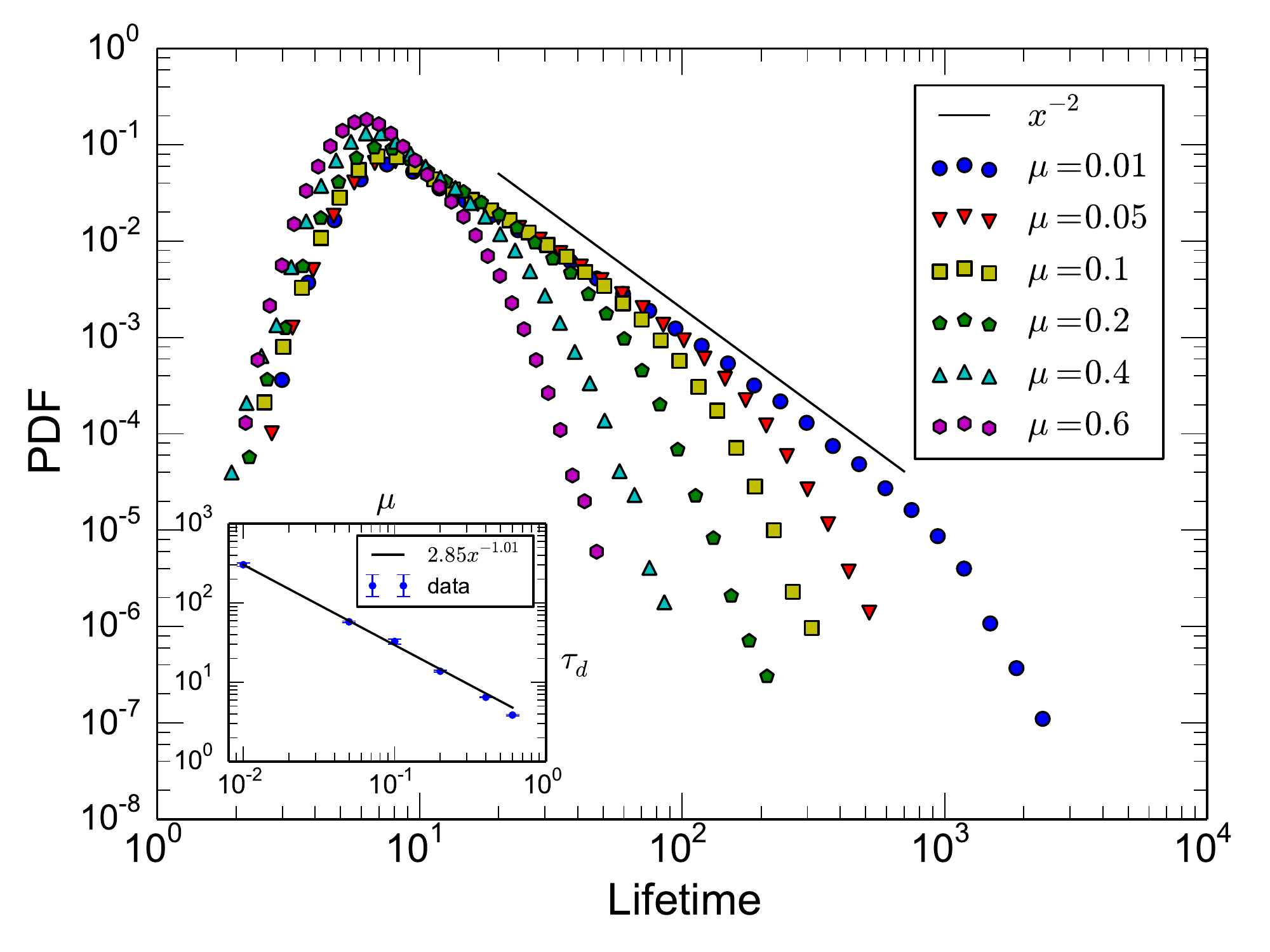}
\caption{
Main: Lifetime PDF for fixed $\alpha=10$ and $N_u=10^3$ obtained by simulating 
the QBC model on an annealed regular network with degree 1 and 
fixed fitness $f=1$. 
The solid straight line is a power-law with exponent $-2$.
Inset: Symbols are the temporal scale $\tau_d$ over which the 
PDF decays, estimated numerically by fitting the exponential tails
in the main. 
The straight line is a power-law fit to the numerical values of $\tau_d$,
confirming that $\tau_d$ is inversely proportional to 
$\mu$ for $f=1$ [see Eq.~(\ref{v0})].}
\label{4.2a}
\end{figure}

This expression of $F(l)$ accounts for the most important feature
observed in simulations: for $l \ll \tau$ (notice that $\tau$
diverges with $N_u$) the distribution decays as a power-law with 
exponent $\eta_l=2$.
Simulations of the QBC model with all memes having fitness $f=1$
agree with this analytical prediction (see Fig.~\ref{4.2a}).
By means of the standard argument connecting the exponents
of power-law tails for scaling variables (see SM) it is possible to relate
$\eta_l$ with the analogous exponent $\eta_p$ for the popularity distribution:
$\eta_l = s(\eta_p-1)+1$, where $p \sim l^s$.
Simulations yield a value close to $s=2$, 
from which $\eta_p=3/2$, in good agreement with simulations (see SM).

\subsection{Pure drift}

The opposite limit for the FP equation (\ref{eq:FP_complete}) 
is the pure drift case, which always holds in the large $N_u$ limit,
as $D_0 \propto N_u^{-1}$, unless $\mu=0$ and $f=1$:
\begin{equation}
  \frac{\partial}{\partial t}\rho(x,t)=\frac{1}{\tau_d}
  \frac{\partial}{\partial x}x\rho(x,t)
\label{eq:FP_my-pure-drift}
\end{equation}
where
\begin{equation}
  \frac{1}{\tau_d} = \frac{1-\beta}{\gamma \alpha}=[1-(1-\mu)f](\alpha+1) .
\label{v0}
\end{equation}
This equation describes a deterministic motion
\begin{equation}
  x(t) = x_{\alpha} e^{-t/\tau_d} \;,
\end{equation}
i.e., the meme position drifts exponentially toward $x=0$;
in other words the systematic drift attracts walkers toward the
absorbing boundary. This introduces an additional exponential
cutoff in the lifetime distribution, which can be globally written as 
\begin{equation}
  F(l) \propto \alpha l^{-2} e^{-l/\tau_d}
  \label{eq:F(l)_full}
\end{equation}
in agreement with simulations (see Fig.~\ref{4.2a}, inset).

\subsection{Average over the fitness}
In the original definition of the QBC model the fitness is a random
variable uniformly distributed between $0$ and $1$. 
Using Eq.~(\ref{eq:F(l)_full}) it is possible to compute the lifetime
distribution also in this case, by averaging over $f$ (see SM) and obtaining, 
in the limit $\mu \rightarrow 0$:
\begin{equation}
  F_{\langle f\rangle}(l) \approx \alpha (\alpha+1) l^{-3} \left[ 1 -e^{-(\alpha+1)l} \right] \; .
\end{equation}
The exponent of the lifetime distribution is then $\eta_l=3$, in
reasonable agreement with Fig.~\ref{fig_annealed}. A similar conclusion
can be drawn for the popularity distribution, predicted to decay as $p^{-2}$.

In summary, by means of a mapping of QBC dynamics onto a 
random-walk description, we have derived expressions for
for the lifetime and popularity distributions, which account 
for the phenomenology observed in numerical simulations.

\section{A macroscopic approach}
\begin{figure}
\includegraphics[width=\columnwidth]{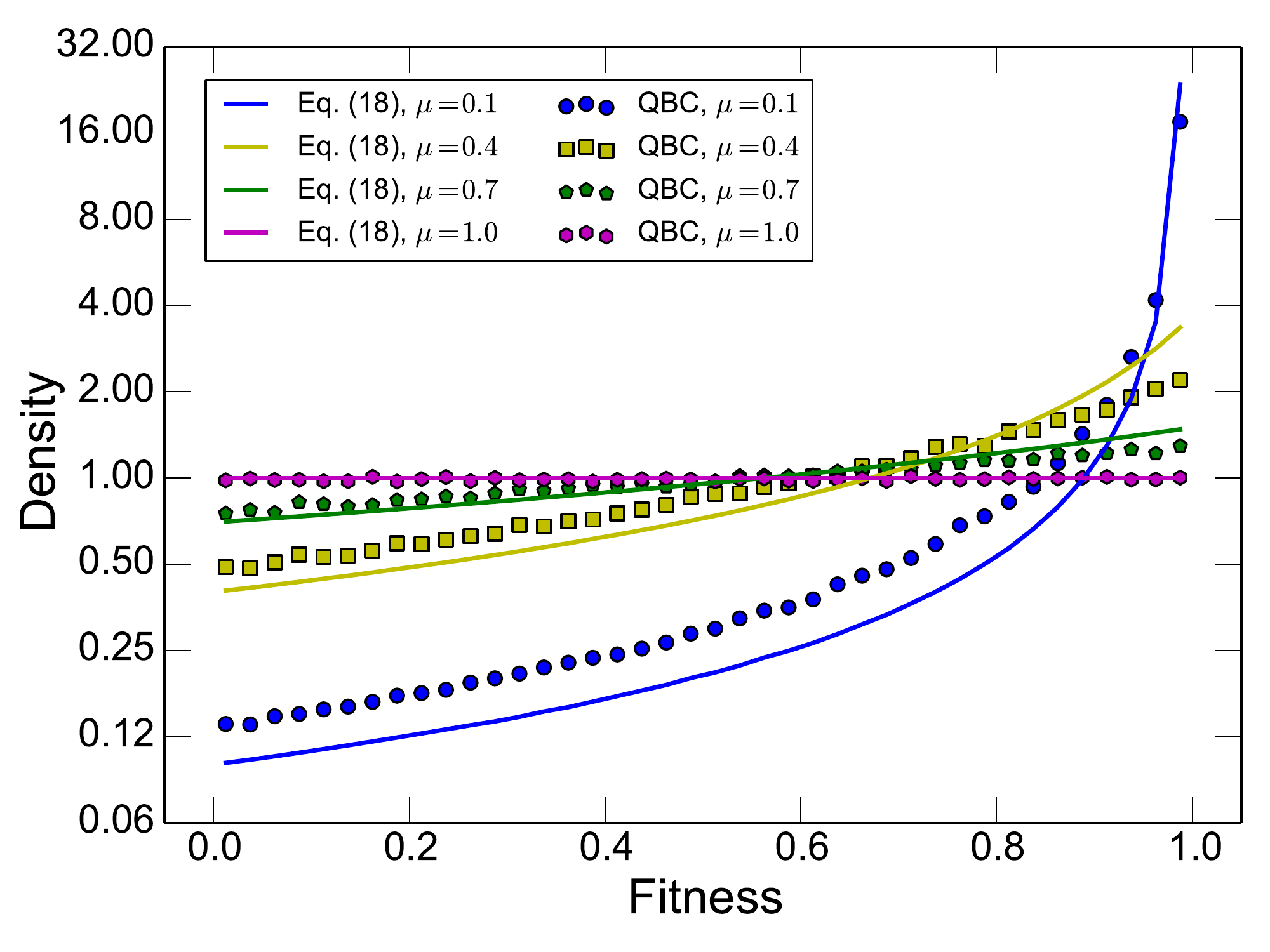}
\caption{Comparison between the average density of memes in the QBC model 
(symbols) and the stationary solutions of Eq.~(\ref{eq:transitions}) (lines)
for different values of $\mu$. $N_u=10^3$, $\alpha=10$, $F=40$. 
Averages are computed over $10^5$ memes.}
\label{densities}	
\end{figure}

The microscopic approach allows to determine the dependence of the
average lifetime on the fitness and hence estimate the average number
$N_f$ of memes with given $f$ in the steady state.
However, the same quantities can be derived much more easily
by a simple approach of mean-field type, focused directly
on the temporal evolution of the $N_f$.
For simplicity we assume that fitness values are discretized in $F$ classes
and, again, that the degree of each agent is 1.
We define $N_f(t)$ as the average number of memes with fitness $f$
present in the system at time $t$. 
This quantity changes over time
because of two possible gain and two possible loss processes.
The creation of a new meme, occurring at rate $\mu$, increases $N_f$ by 1
with a probability $1/F$ (if the created meme has exactly fitness $f$), but
it may also reduce $N_f$ by 1 if the agent creating the new meme forgets
a meme of fitness $f$. This last event occurs with probability 
$N_f/(N_u\alpha)$.
The transmission of an existing meme, occurring at rate $1-\mu$,
increases $N_f$ if the transmitting agent has a meme with fitness $f$
in her feed (probability proportional to $N_f$) and the meme is selected
(probability proportional to $f$). Overall the normalized
probability of the event is $f N_f/[\sum_{f'} f' N_{f'}]$.
Finally also the transmission event may lead to an agent forgetting a meme
with fitness $f$ with probability $N_f/(N_u\alpha)$. The temporal
evolution of the $N_f$ is then given by the set of coupled equations
\begin{equation}
\!\dot{N}_f(t)\!=\!\!\mu \!\left[\frac{1}{F} - \frac{N_f(t)}{N} \right] + (1-\mu)\!\left[ \frac{f N_f(t)}{\sum_{f'} f' N_{f'}(t)} - \frac{N_f(t)}{N}\! \right]\!.
\label{eq:transitions}
\end{equation}
which conserves the total number $\sum_{f'} N_{f'}$.
Straightforward numerical integration of Eq.~(\ref{eq:transitions}) allows to
determine the stationary values of the $N_f$ and hence of the 
densities $n_f=N_f/(N_u\alpha/F)$, where $N_u\alpha/F$ is the average number of
memes with fitness $f$ if all $F$ classes were populated uniformly.
The comparison with the outcome of numerical simulations 
(see Fig.~\ref{densities})
confirms a satisfactory agreement.

\section{Conclusions}

In this paper we have studied the model for
information diffusion recently introduced in Ref.~\cite{Qiu2017}.
We have been able to derive analytically the lifetime distribution
and other properties for a simplified version of the dynamics, which
reproduces the phenomenology of the original model.

Our treatment of the QBC model allows to understand how broad tails in
the lifetime and popularity distributions, observed empirically, arise.  
A power-law distribution with an 
exponent $\eta_p \approx 2$ is in agreement with the observations
of Ref.~\cite{Qiu2017}, where hashtags are used to identify Twitter 
memes.
On the other hand, other studies using hashtags give quite 
different results from the QBC model predictions. 
In Ref.~\cite{Weng2012} a power-law decay for meme lifetime has been
observed, with an exponent $\eta_l \approx 2.5$.
This value is not far
but distinct from the value $\eta_l$ predicted by the QBC model in the
case of uniformly distributed fitness. Moreover, the strong
correlation between meme lifetime and popularity (see SM) is not
observed in Twitter data~\cite{Banos2013,Gleeson2016}, even if
proxies different from hashtags are used to identify 
memes~\cite{GonzalezBailon2011}. 
A stringent empirical validation of models of online information 
spreading  is itself a difficult task because of the apparent lack of
universality. Referring to Twitter data, the identification of memes
as URLs leads to a lognormal distribution of
popularity~\cite{Lerman2010}, the analysis of retweet cascades leads
to a size distribution with exponent $\eta_s \approx 2.3$~\cite{Wegrzycki2017} 
with possibly an exponential cutoff~\cite{Vosoughi1146} and reply trees 
give $\eta_s \approx 4$~\cite{Nishi2016}. 
Looking at other data sources, the landscape is
even more varied: the popularity distribution, estimated from Facebook data,
exhibits a power-law deacy with exponent $\eta_p \approx 2.1$~\cite{Cheng2014}, 
while popularity-lifetime correlations are shown to be different 
between Digg and Youtube data~\cite{szabo2010predicting}. 
One could easily change, within the QBC model, the fitness distribution 
to achieve a better agreement with these observations.  
In any case it is clear that the QBC model is a
gross oversimplification of the real meme diffusion process in online
social media.  To make the QBC dynamics less unrealistic several
hypotheses underlying the present version of model could be lifted.
Some of them, such as a nonuniform fitness distribution or a nonlinear
dependence on $f$ of the probability of selecting a meme, can be
easily treated within the present analytical approach.  Other
fundamental generalizations, such as agent-dependent values of
$\alpha$ and $\mu$ or heterogeneous rates of individual activation,
can be investigated by means of straightforward numerical simulations.
One of the ingredients adding realism to the QBC dynamics is the
consideration of agents that do not accept in their feeds (and thus do
not spread further) memes they have already seen in the past.  The
effect of this long-term memory is briefly discussed in the SM, but
the main result is the change of the popularity and lifetime
distributions, that lose their power-law tail.  At a more general level,
one of the weak points of the QBC model is its insensitivity with
respect to changes of the contact pattern topology.  While this
feature allows our mean-field approach to be successful, empirical
data contradict this result: one of the main pieces of evidence is the 
existence of influential spreaders, i.e. users which, because of their 
position in the social network have a disproportionate effect on meme
dynamics~\cite{Bakshy2011,Banos2013,borge2012locating}. The investigation
of increasingly sophisticated models for information spreading and the
comparison with the ever larger body of empirical data available
remains a challenging avenue for future research.

\bibliography{qbc}
\clearpage
\pagenumbering{gobble}

\begin{widetext}
\vspace*{-2.2cm}
\hspace*{-1.8cm}\includegraphics{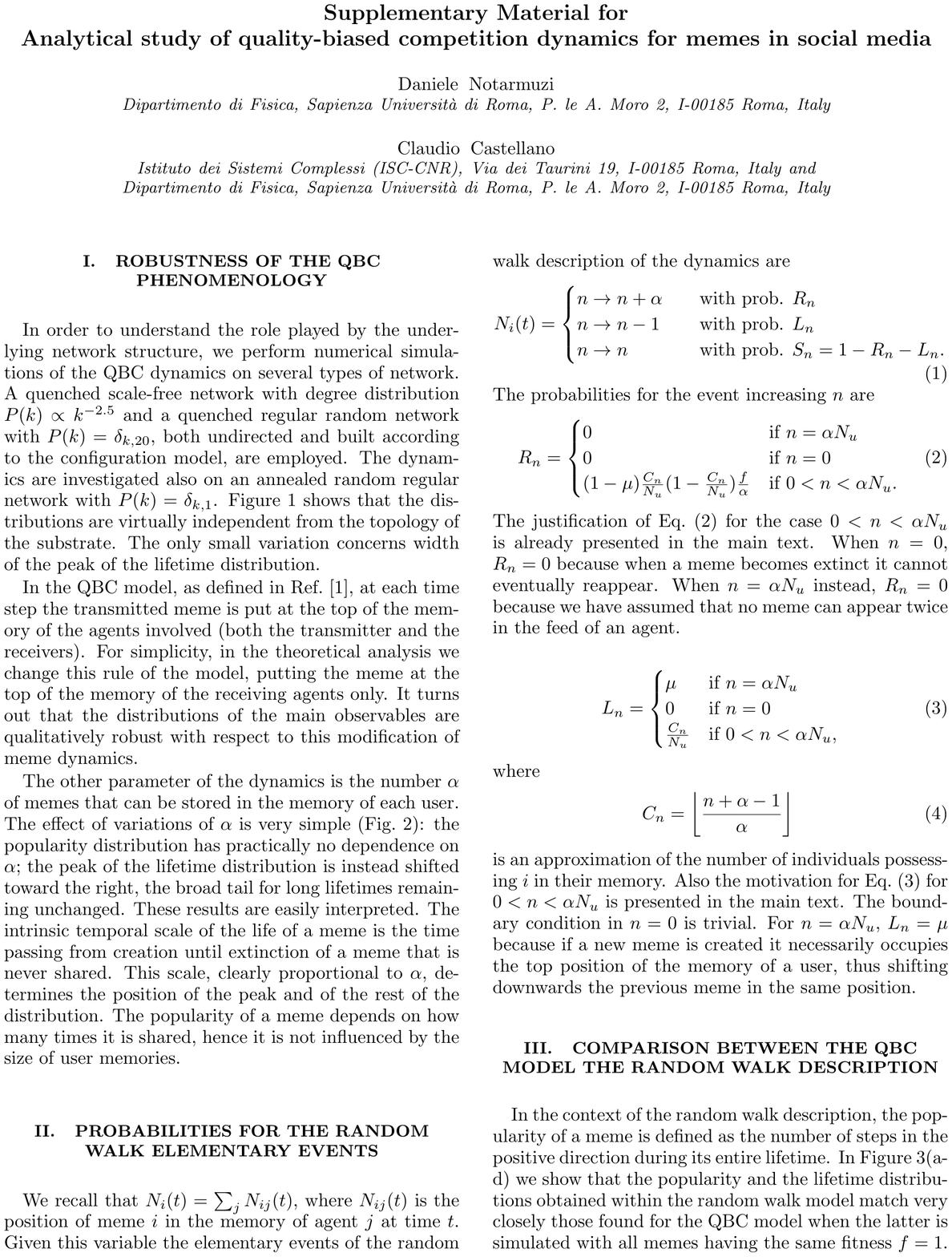}\\

\vspace*{-2.2cm}
\hspace*{-1.8cm}\includegraphics{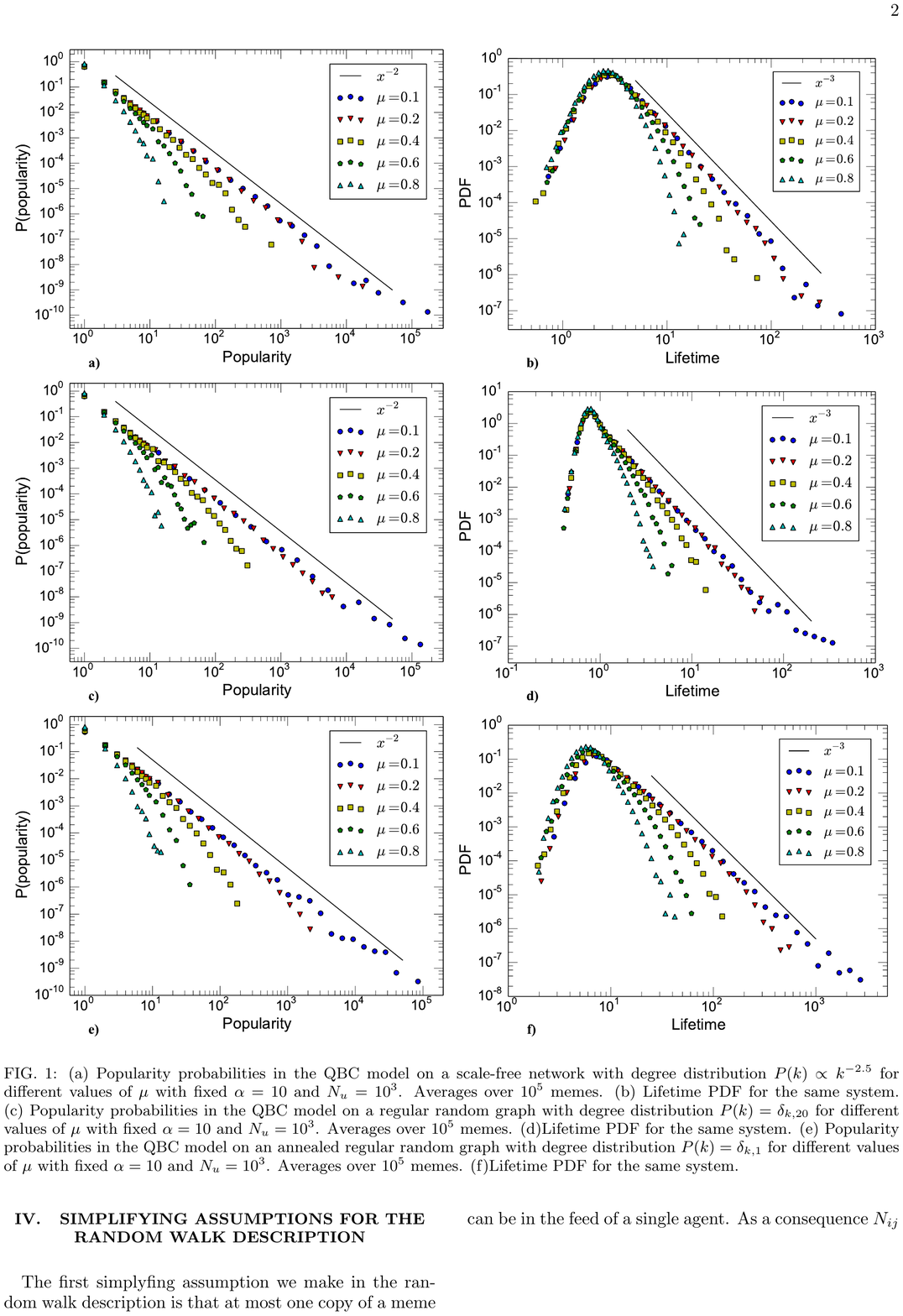}\\

\vspace*{-2.2cm}
\hspace*{-1.8cm}\includegraphics{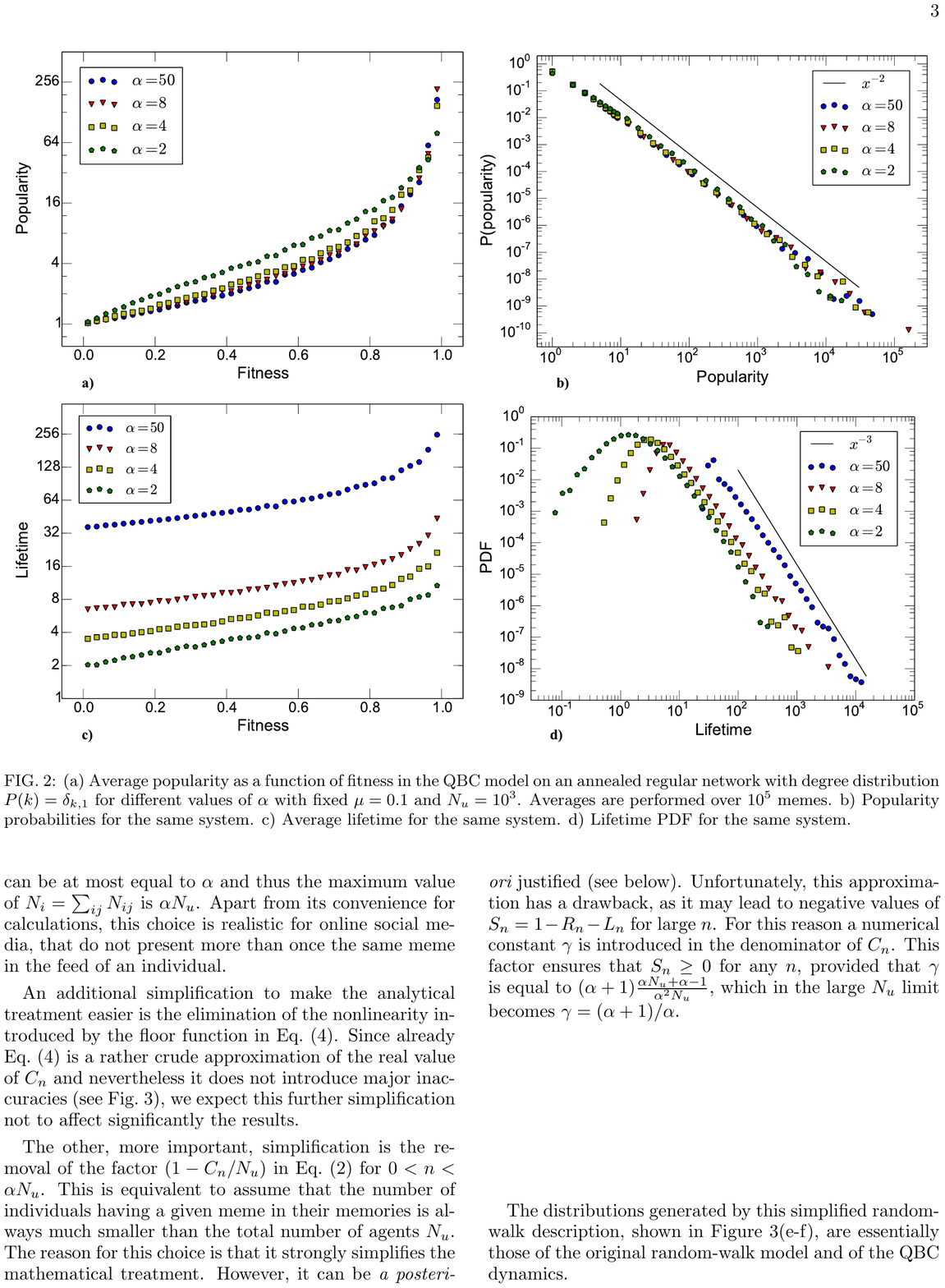}\\

\vspace*{-2.2cm}
\hspace*{-1.8cm}\includegraphics{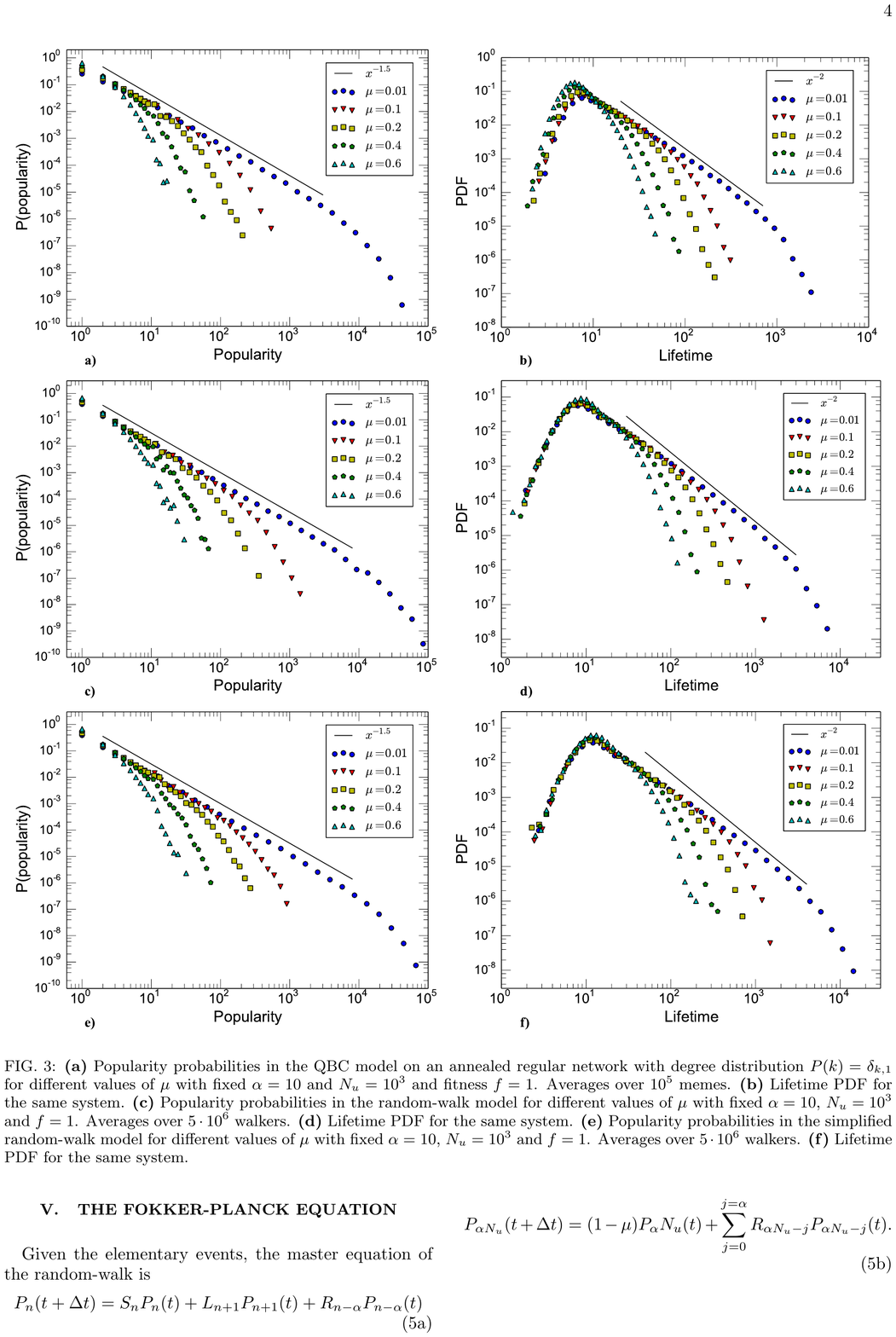}\\

\vspace*{-2.2cm}
\hspace*{-1.8cm}\includegraphics{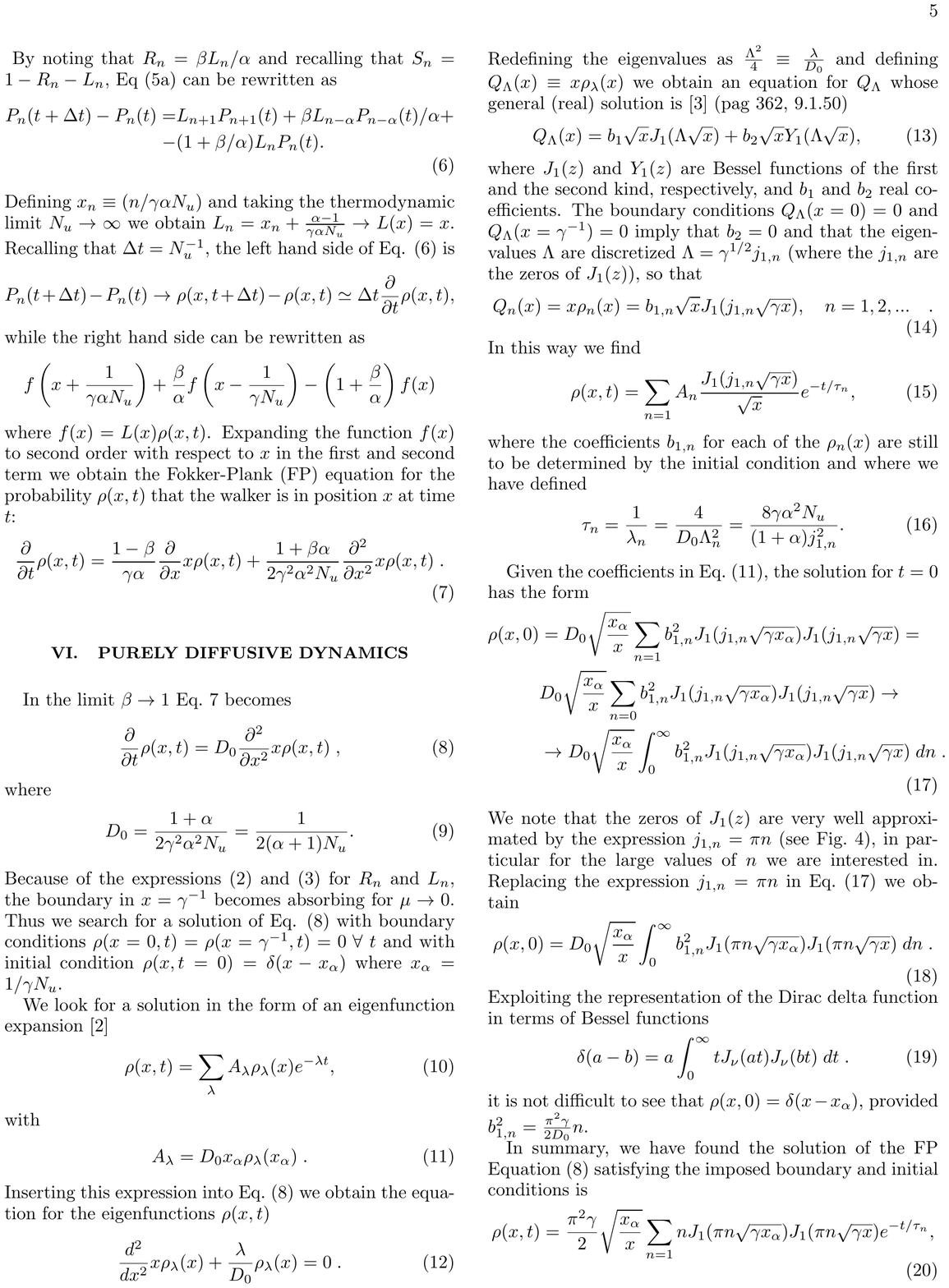}\\

\vspace*{-2.2cm}
\hspace*{-1.8cm}\includegraphics{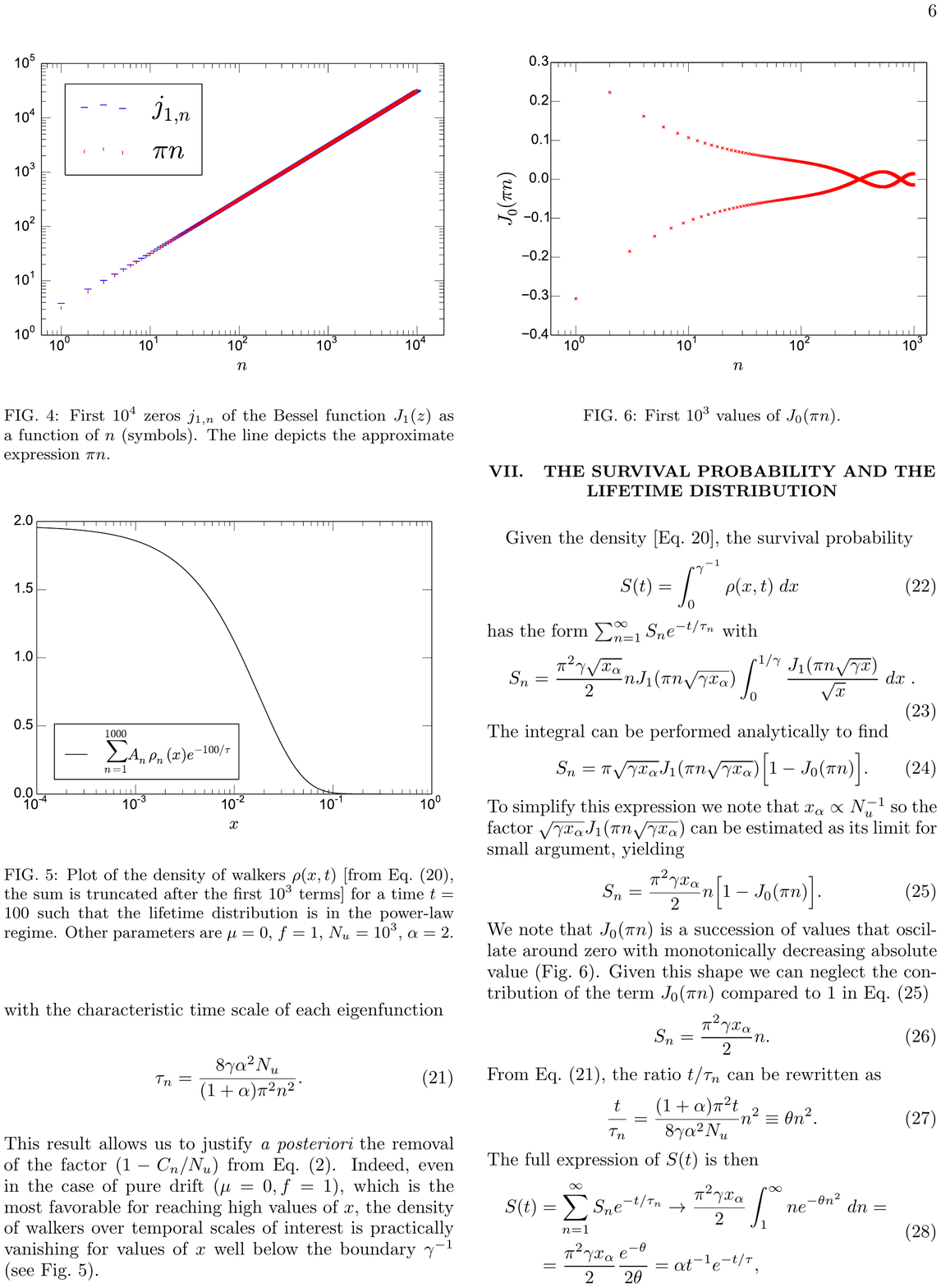}\\

\vspace*{-2.2cm}
\hspace*{-1.8cm}\includegraphics{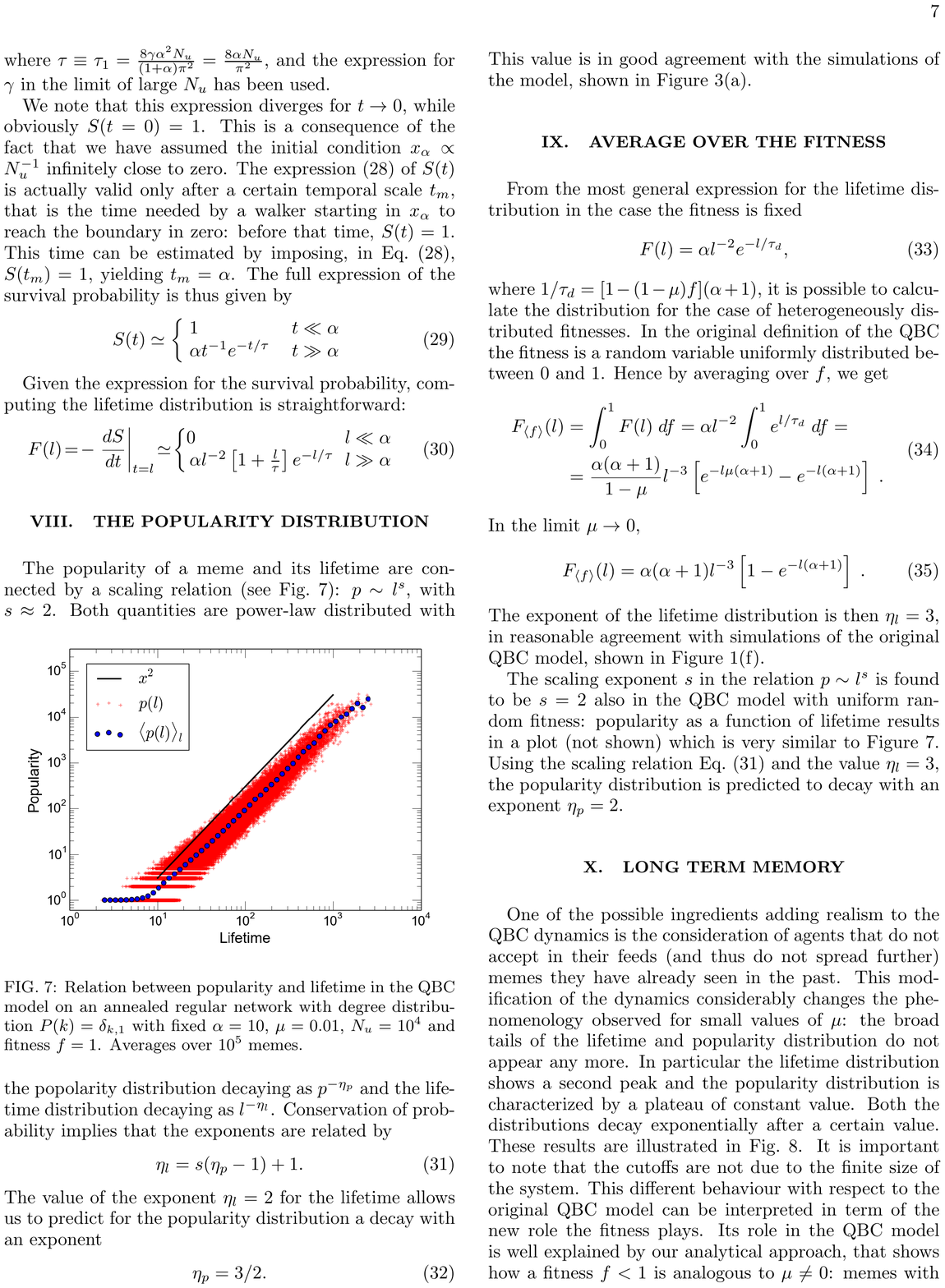}\\

\vspace*{-2.2cm}
\hspace*{-1.8cm}\includegraphics{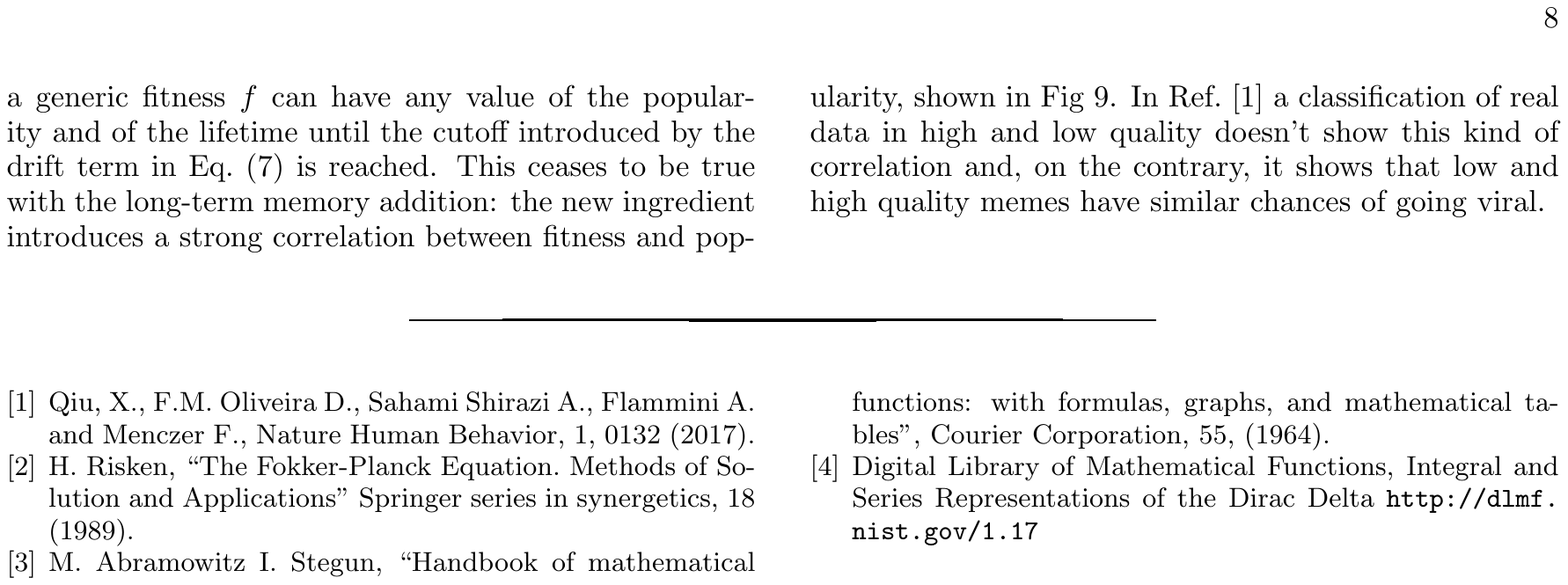}\\

\vspace*{-2.2cm}
\hspace*{-1.8cm}\includegraphics{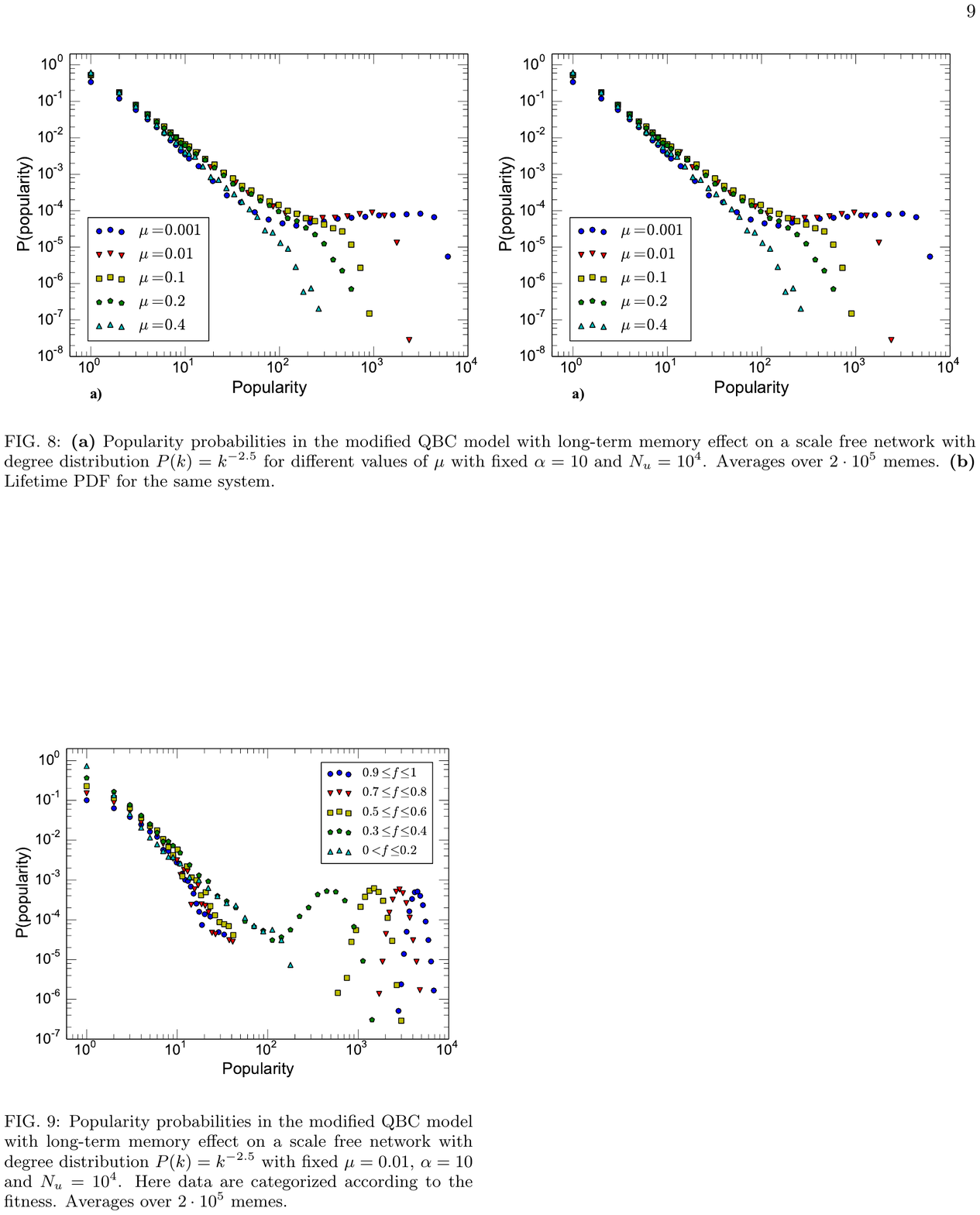}
\end{widetext}

\end{document}